\documentclass[twocolumn,aps,prl,floatfix,longbibliography,groupedaddress]{revtex4-1}
\usepackage{amsmath,bbm}
\usepackage{graphicx}
\usepackage{multirow}
\usepackage{hyperref}
\usepackage{amsfonts}
\usepackage{amsbsy}
\usepackage{xcolor}
\usepackage{soul}
\usepackage{graphicx}
\usepackage{mathtools}

\renewcommand{\vec}[1]{\mathbf{#1}}

\newcommand{\e}{\mathcal{E}}
\newcommand{\Pc}{\mathcal{P}}
\newcommand{\cbE}{\boldsymbol{\mathbf{\cal E}}}

\newcommand{\ket}[1]{\ensuremath{\left\vert #1 \right\rangle}}
\newcommand{\unitvec}[1]{\hat{\mathbf{{#1}}}}
\renewcommand{\vec}[1]{\mathbf{#1}}

\begin{document}
\title{Radiative toroidal dipole and anapole excitations in collectively responding arrays of atoms}

\date{\today}
\author{K.~E.~Ballantine}
\email{k.ballantine@lancaster.ac.uk}
\author{J.~Ruostekoski}

\email{j.ruostekoski@lancaster.ac.uk}

\affiliation{Department of Physics, Lancaster University, Lancaster, LA1 4YB, United Kingdom}

\begin{abstract}
A toroidal dipole represents an often overlooked electromagnetic excitation distinct from the standard electric and magnetic multipole expansion. We show how a simple arrangement of strongly radiatively coupled atoms
can be used to synthesize a toroidal dipole where the toroidal topology is generated by radiative transitions forming an effective poloidal electric current wound around a torus. We extend the protocol for methods to
prepare a delocalized collective excitation mode consisting of a synthetic lattice of such toroidal dipoles and a non-radiating, yet oscillating charge-current configuration, dynamic anapole, for which the far-field radiation of 
a toroidal dipole is identically canceled by an electric dipole.

\end{abstract}

\maketitle

The concept of electric and magnetic multipoles vastly simplifies the study of light-matter interaction, allowing the decomposition both of the scattered light and of the charge and current sources~\cite{Jackson,Raab05}. While the far-field radiation can be fully described by the familiar transverse-electric and -magnetic multipoles,  the full characterization of the current requires, however, an additional series that is independent of electric and magnetic multipoles: dynamic toroidal multipoles~\cite{Papasimakis16,Talebi18,Dubovik90,Vrejoiu02,Dubovik90}. 
These are extensions of static toroidal dipoles~\cite{Zeldovich58} that have been studied in nuclear, atomic, and solid-state physics, e.g., in the context of parity violations in electroweak interactions~\cite{Dmitriev2004243,Flambaum,Wood97} and in multiferroics~\cite{Schmid}.
Often obscured and neglected in comparison to electric and magnetic multipoles due to its weakness, the toroidal dipole can have an important response to electromagnetic fields in systems of toroidal geometry~\cite{Savinov14}. 
Dynamic toroidal dipoles are actively studied in artificial metamaterials that utilize such designs, with responses varying from microwave to optical part of the spectrum~\cite{Kaelberer10,Ogut12,Dong12b,Fan13,Basharin15,Watson2016}.
Crucially, an electric dipole together with a toroidal dipole can form a non-radiating dynamic anapole~\cite{Afanasiev95,Savinov19,Fedotov13,Miroshnichenko15}, where the far-field emission pattern from both dipoles interferes destructively, so the net emission is zero.

Light can mediate strong resonance interactions between closely-spaced ideal emitters and an especially pristine system exhibiting cooperative optical response is that of regular planar arrays of cold atoms with unit occupancy~\cite{Jenkins2012a,Perczel2017a,Bettles_lattice,Facchinetti16,Yoo16,Asenjo-Garcia2017a,Ritsch_subr,
Mkhitaryan18,Shahmoon,Guimond2019,Javanainen19,bettles2019,Qu19,Ballantine20ant}.
Subradiant linewidth narrowing of transmitted light was now observed in such a system formed by an optical lattice~\cite{Rui2020}. In the experiment the whole array was collectively responding to light with the atomic dipoles oscillating in phase. Furthermore, the lattice potentials can be engineered~\cite{Wang18}, and a great flexibility of optical transitions is provided by atoms such as Sr and Yb~\cite{Olmos13}.
Also several other experimental approaches exist to trap and arrange atoms with single-site control~\cite{Lester15,Xia15,Endres16,Barredo16,Kim16,Cooper18,Glicenstein20}.

Here we propose how to harness strong light-mediated interactions between atoms to engineer collective radiative excitations that synthesize effective dynamic toroidal dipole and non-radiating anapole moments, even though individual
atoms only exhibit electric dipole transitions. The method is based on simple arrangements of atoms, where the toroidal topology is generated by radiative transitions forming an effective poloidal electric current wound around a torus, such that an induced magnetization forms a closed circulation inside the torus. The toroidal and anapole modes can be excited by radially polarized incident light and, in the case of the anapole, with a focusing
lens. The resulting anapole excitation shows a sharp drop in the far-field dipole radiation, despite having a large collective electronic excitation of the atoms associated with both electric and toroidal dipole modes. Such a configuration 
represents stored excitation energy without radiation that is fundamentally different, e.g., from subradiance~\cite{Dicke54}. We extend the general principle to larger systems, and show in sizable arrays how this collective behavior of atoms allows us to engineer a delocalized collective radiative excitation eigenmode consisting of an effective periodic lattice of toroidal dipoles.
Such an array is then demonstrated to exhibit collective subradiance, with the narrow resonance
line sensitively depending on the lattice spacing and manifesting itself as a Fano resonance in the
coherently transmitted light.

Utilizing cold atoms as a platform for exploring toroidal excitation topology has several advantages, as it naturally allows for the toroidal response at optical frequencies, with emitters much smaller than a resonant wavelength, and avoids dissipative losses present in plasmonics or circuit resonators forming metamaterials.
Moreover, the atomic arrangement can form a genuine quantum system, and our analysis is valid also in the single photon quantum limit.

\begin{figure}
\centering
		\includegraphics[width=1\columnwidth]{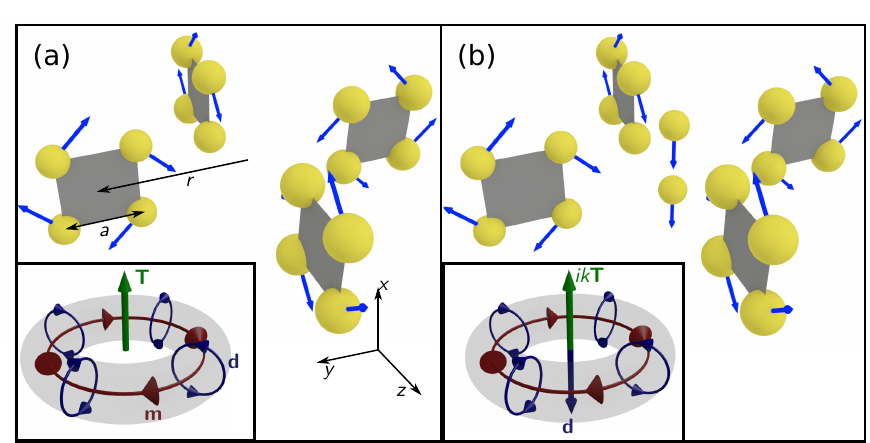}
    \vspace*{-12pt}
		\caption{
		(a) Geometry of a toroidal dipole unit cell consisting of a number of four-atom squares of width $a$, arranged in a circle of radius $r$, with atomic dipoles indicated by the arrows. A circulating electric polarization on the surface of a torus leads to magnetic dipoles forming a closed loop inside a torus (inset).
These in turn contribute to a toroidal dipole moment through the center of the entire unit cell. (b) Geometry of an anapole unit cell. Adding two atoms to the center with induced dipole moments in the $x$ direction generates an electric dipole which  destructively interferes with the toroidal dipole (inset).  
		}
		\label{fig:geometry}
\end{figure}

\begin{figure}[ht]
\centering
		\includegraphics[width=1\columnwidth]{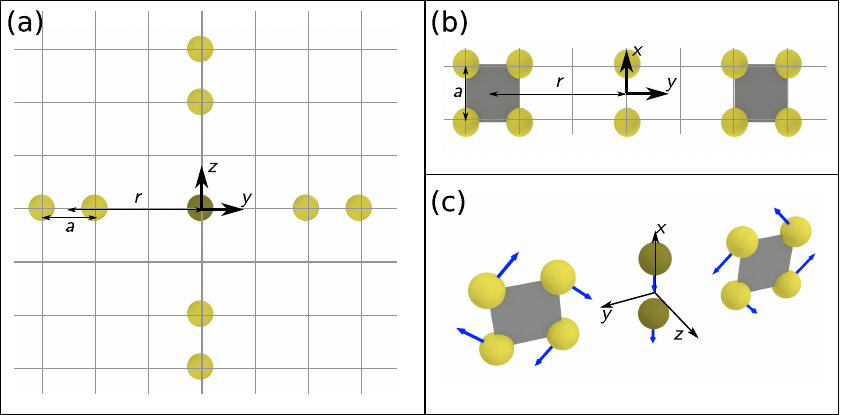}
    \vspace*{-12pt}
		\caption{
		Projections of geometry of toroidal dipole and anapole unit cells in (a) The $yz$ and (b) the $xy$ plane. Shaded atoms at $y,z=0$ are absent for toroidal dipole unit cell but present for anapole unit cell. (c) Alternative structure for realizing a toroidal dipole or anapole excitation with all atoms in the $xy$ plane the response of which is shown in Fig.~S3~\cite{SOM}. 
		}
		\label{fig:geo_supp}
\end{figure}

To demonstrate the formation of an effective toroidal dipole and anapole in an atomic ensemble, we briefly describe the radiative coupling between cold atoms. For simplicity of presentation, we analyze coherently driven case,
but the formalism is also valid in the quantum regime of single photon excitations~\cite{SOM}. 

We consider atoms at fixed positions, with a $J=0\rightarrow J^\prime=1$ transition, and assume a controllable Zeeman splitting of the $J^\prime=1$ levels, generated, e.g., by a periodic optical pattern of ac Stark shifts~\cite{Ballantine20}. The dipole moment of atom $j$ is 
$\vec{d}_j = \mathcal{D}\sum_j \Pc_{\sigma}^{(j)}\unitvec{e}_\sigma$,
where $\mathcal{D}$ denotes the reduced dipole matrix element, and $\Pc_{\sigma}^{(j)}$ and $\unitvec{e}_\sigma$ the polarization amplitude and unit vector associated with the $\ket{J=0,m=0}\rightarrow\ket{J^\prime=1,m=\sigma}$ transition, respectively. The collective response of the atoms in the limit of low light intensity~\cite{Morice1995a,Ruostekoski1997a,Javanainen1999a,Sokolov2011,Jenkins2012a,Lee16}  then follows from
$\dot{\vec{b}}  = i\mathcal{H}\vec{b}+\vec{f}$,
where $\vec{b}_{3j-1+\sigma}=\Pc_{\sigma}^{(j)}$ and the driving $\vec{f}_{3j-1+\sigma}=i(\xi/\mathcal{D})\unitvec{e}_\sigma^\ast\cdot\epsilon_0\cbE(\vec{r}_j)$, with the incident light field of amplitude $\cbE(\vec{r})=\mathcal{E}_0\unitvec{e}_{\rm in} \exp{(ikx})$~\cite{foot}, polarization $\unitvec{e}_{\rm in}$, and frequency $\omega=kc$. Here $\xi=6\pi\gamma/k^3$ depends on the single-atom linewidth $\gamma=\mathcal{D}^2k^3/(6\pi\epsilon_0\hbar)$. The matrix $\mathcal{H}$ describes interactions between different atoms due to multiple scattering of light, with $\mathcal{H}_{3j-1+\sigma,3k-1+\sigma^\prime} = \xi \unitvec{e}_\sigma^\ast\cdot\mathsf{G}(\vec{r}_j-\vec{r}_k)\unitvec{e}_{\sigma^\prime}$ for $(j,\sigma)\neq (k,\sigma^\prime)$, where the dipole radiation kernel $\mathsf{G}(\vec{r})$ gives the field $\epsilon_0\vec{E}_s^{(j)}(\vec{r})=\mathsf{G}(\vec{r}-\vec{r}_j)\vec{d}_j$ from a dipole moment $\vec{d_j}$ at $\vec{r}_j$~\cite{Jackson}. The diagonal element $\mathcal{H}_{3j+\sigma-1,3j+\sigma-1}=\Delta_{\sigma}^{(j)}+i\gamma$, where $\Delta_\sigma^{(j)}=\delta_\sigma^{(j)}+\Delta$ consists of an overall laser detuning $\Delta=\omega-\omega_0$ from the single-atom resonance $\omega_0$, plus a relative shift $\delta_\sigma^{(j)}$ of each level. The dynamics follows from eigenvectors $\vec{v}_n$ and eigenvalues $\delta_n+i\gamma_n$ of $\mathcal{H}$ giving the collective level shifts $\delta_n$ and linewidths $\gamma_n$~\cite{Jenkins_long16}. 

The limit of low light intensity corresponds to linear regime of oscillating atomic dipoles. The analogous quantum limit is that of a single photon that experiences no nonlinear interactions, as at minimum, two simultaneous photons
are required for interactions. Regarding one-body expectation values,
the equations of motion for the dipole amplitudes $\Pc_{\sigma}^{(j)}$ are indeed precisely the same~\cite{SVI10} as those for single-photon excitation amplitudes that are radiatively coupled between the atoms by $\mathcal{H}$~\cite{SOM}.

To illustrate the role of toroidal multipoles, we consider the far-field scattered light from a radiation source  decomposed into vector spherical harmonics~\cite{Jackson}, 
\begin{equation}
\vec{E}_s^{(j)} = \sum_{l=0}^\infty\sum_{m=-l}^{l}\left(\alpha_{\mathrm{E},lm}^{(j)}\vec{\Psi}_{lm}+\alpha_{\mathrm{B},lm}^{(j)}\vec{\Phi}_{lm}\right),
\label{eq:harmonics} 
\end{equation}
that allows us to represent it as light originated from a set of multipole emitters at the origin, with $l=1$ representing dipoles, $l=2$ quadrupoles, etc. However, while the magnetic coefficients $\alpha_{\mathrm{M},lm}$ are due to magnetic multipole sources with transverse current $\vec{r}\times\vec{J}\neq 0$, the electric coefficients $\alpha_{\mathrm{E},lm}$ can arise from two different types of polarization; electric and toroidal multipoles.
These contributions can be calculated directly from the induced polarizations. Taking atom $j$ to be fixed at position $\vec{r}_j$, the induced displacement current density is $J_\sigma(\vec{r})=-i\omega\mathcal{D}\sum_j\Pc_\sigma^{(j)}\delta(\vec{r}-\vec{r_j})$. Inserting this in the standard multipole decomposition for an arbitrary distribution of currents~\cite{Vrejoiu02} gives  for the total electric and  magnetic dipoles $\vec{d} = \sum_j \vec{d}_j$ and $\vec{m} = -(ik/2)\sum_j(\vec{r}_j\times\vec{d}_j)$, respectively, and for the  toroidal dipole  
\begin{align}
\label{eq:t}
\vec{T} &= -\frac{ik}{10} \sum_j \left[\left(\vec{r}_j\cdot\vec{d}_j\right)\vec{r}_j-2\vec{r}_j^2\vec{d}_j\right].
\end{align}
The magnitude of the far-field electric dipole component, $|\alpha_{\mathrm{E},1}|\equiv [\sum_m|\alpha_{\mathrm{E},1m}|^2]^{1/2}\propto k^2|\vec{p}|/(4\pi\epsilon_0)$ then depends on the combination $\vec{p}=\vec{d}+ik\vec{T}$~\cite{Miroshnichenko15}. We have checked that in our numerics corrections beyond the long-wavelength approximation of Eq.~\eqref{eq:harmonics} are negligible.

\begin{figure}
\centering
		\includegraphics[width=1\columnwidth]{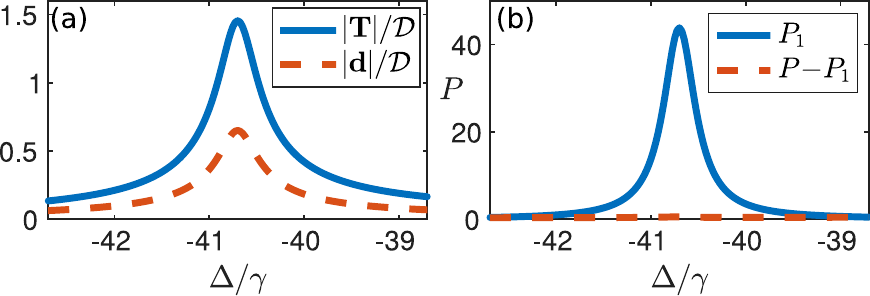}
    \vspace*{-12pt}
		\caption{Excitation of a toroidal dipole $\vec{T}$ as a multipole decomposition of (a) atomic dipoles [in dimensionless units of $\mathcal{D}\mathcal{E}_0/(\hbar\gamma)$]. (b) far-field radiated power consisting of dipole contribution $P_1$ and the remaining power of all other contributions (in units of $I_\mathrm{in}/k^2$, where $I_\mathrm{in}=2c\epsilon_0|\mathcal{E}_0|^2$ is the incoming intensity), as a function of the laser detuning from the atomic resonance, with $r=0.2\lambda$ and $a=0.08\lambda$, as defined in Fig.~\ref{fig:geometry}(a). 
		}
		\label{fig:toroidal}
\end{figure}

We now turn to the design and preparation of a collective toroidal dipole. Even for atoms exhibiting electric dipole transitions, their collective excitation eigenmodes can be utilized in synthesizing radiative excitations, e.g., with magnetic properties~\cite{Ballantine20,Alaee20}. The toroidal dipole, as illustrated in the inset of Fig.~\ref{fig:geometry}(a), consists of a poloidal electric current wound around a torus, such that magnetic dipoles form a closed loop, reminiscent of vortex current, pointing along a ring around the center of the torus. We approximate this geometry using squares of four atoms [see Fig.~\ref{fig:geometry}(a)]. This is possible, since an isolated square has a collective excitation  eigenmode with the dipoles oriented tangentially to the center of the square~\cite{Ballantine20}. While electric dipoles of the atoms average to zero on each square, they generate a magnetic dipole moment normal to the plane of the square. Arranging several of these squares in a circle, with each aligned perpendicular to the circumference, leads to the magnetic dipole moments winding around the center, as illustrated in the inset, creating a toroidal dipole pointing in the $x$ direction. The projections of this geometry in the $yz$ and $xy$ planes are shown in Fig.~\ref{fig:geo_supp}(a,b). A general choice of parameters could be realized with independent optical tweezers. However, in the case that $r=(n+1/2) a$ for integer $n$, the ensemble could also be formed by selectively populating sites on a bilayer square lattice, as indicated by the grey grid.

We demonstrate this by an example calculation of four such squares, with $r=0.2\lambda$ and $a=0.08\lambda$ [Fig.~\ref{fig:geometry}(a)], resulting in altogether 48 collective excitation eigenmodes. This corresponds to a bilayer configuration of atoms, confined to selective sites of a square lattice with lattice constant $0.08\lambda$. We find a collective eigenmode exhibiting a strong toroidal dipole, with only a weak radiative coupling due to subradiant resonance linewidth $\upsilon=0.2\gamma$. The scattered light from this eigenmode is dominated by $|\alpha_{\mathrm{E},1}|^2$ with $>99\%$ of the radiated power coming from this contribution, which can be decomposed locally into toroidal and electric dipoles with $|{\bf T}|/|{\bf d}|=2.2$. At larger lattice spacings, the toroidal dipole contribution can get even more dominant.

To excite the toroidal dipole mode, we consider a plane wave propagating in the $x$ direction. The toroidal symmetry of the mode inhibits coupling to a drive field with uniform linear polarization.
Instead, the symmetry can be matched by radial polarization  $\unitvec{e}_{\rm in}=\unitvec{e}_\rho $, where $\unitvec{e}_\rho$ points outwards in the $yz$ plane from the center of the toroidal dipole. The multipole decomposition of the local excitation in Fig.~\ref{fig:toroidal}(a) displays a strong response of the toroidal dipole, as well as a weaker electric dipole response. Figure~\ref{fig:toroidal}(b) shows the decomposition of the far-field power $P=2c\epsilon_0\int|\vec{E}|^2\,\mathrm{d}A$ integrated over a closed surface into the dominant dipole component $P_1\propto|\alpha_{\mathrm{E},1}|^2$, which does not distinguish between contributions from $\vec{d}$ and $\vec{T}$, as well as the remaining sum of all other contributions. At the toroidal dipole resonance $\Delta=-40.7\gamma$ the occupation of the collective eigenmode is $\approx 99\%$~\cite{SOM}.

\begin{figure}
\centering
		\includegraphics[width=1\columnwidth]{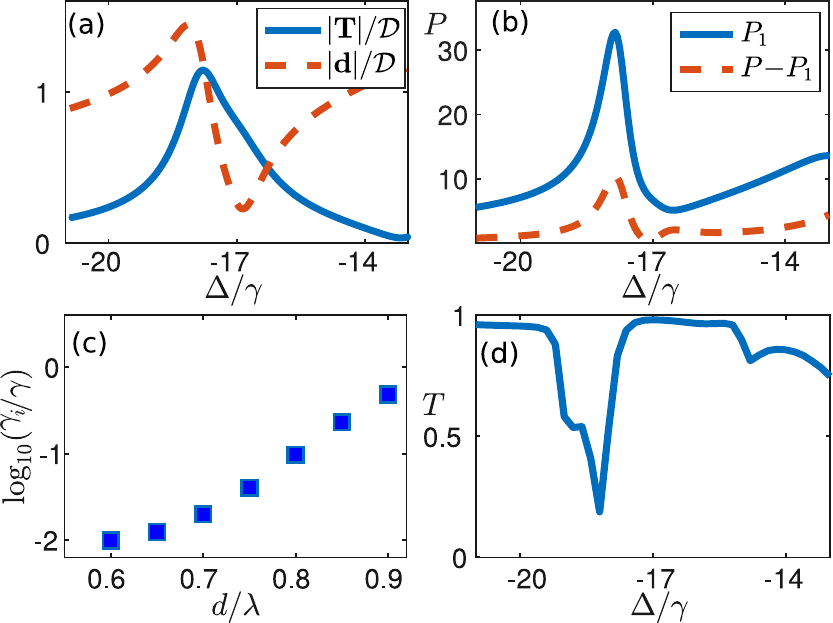}
    \vspace*{-12pt}
		\caption{Collective excitation of a $12\times12$ square array of toroidal dipole unit cells with spacing $d=0.85\lambda$ ($a=0.1\lambda$, $r=0.2\lambda$). (a) Multipole decomposition of
		atomic dipoles [in units of $\mathcal{D}\mathcal{E}_0/(\hbar\gamma)$] for a single central unit cell, excited by linearly polarized light; (b) decomposition of the dipole contribution to the far-field radiated power, and the sum of all other contributions (in units of $I_\mathrm{in}/k^2$); (c) collective linewidth $\upsilon$ of the uniform toroidal dipole eigenmode as a function of $d$; (d) coherent transmission $T=|\vec{E}|^2/|\e_0|^2$ of light through the array, where $\vec{E}$ is the total field amplitude.
		}
		\label{fig:toroidal_lin}
\end{figure}

Here we take four squares, distributed evenly on a ring around the center, to form the toroidal dipole moment, but similar results can be achieved with a minimum of only two. As illustrated in Fig.~\ref{fig:geo_supp}(c), with two squares centered at, e.g.,\ $\pm r \unitvec{y}$ having opposite chirality dipole orientation a toroidal dipole moment can also be achieved while all atoms lie in the single  $xy$ plane~\cite{SOM}.

We next consider a planar square lattice in the $yz$ plane with each unit cell as in  Fig.~\ref{fig:geometry}(a). 
Due to radiative interactions, for a subwavelength-spaced lattice, the entire system responds as a coherent, collective entity, with delocalized collective excitation eigenmodes extending over the array. In particular, there is a collective eigenmode which corresponds to a uniform excitation of a toroidal dipole at each site. However, this mode cannot be excited by radially polarized light as it would require the symmetry to be broken around the center of each individual unit cell. Instead, we use uniform linear polarization, with  $\unitvec{e}_{\rm in}=\unitvec{e}_y$, but vary the atomic level shifts within each atom of the unit cell independently that are then repeated across the array on each unit cell.  We numerically optimize the toroidal dipole moment on a single unit cell to find these level shifts numerically.

The corresponding toroidal and electric dipole excitations are shown in Fig.~\ref{fig:toroidal_lin}(a). Despite the presence of the electric dipole, the toroidal dipole is the dominant component at $\Delta=-17\gamma$ where the ratio $|{\bf T}|/|{\bf d}|=3.3$ is at its maximum. The dipole radiation is compared to the intensity of all other contributions to the scattered light in Fig.~\ref{fig:toroidal_lin}(b), showing that all other modes are also suppressed at this detuning.

This excitation closely corresponds to an eigenmode, delocalized across the entire array, consisting of a repetition of the poloidal dipole excitation on each unit cell, and forming an effective lattice of coherently oscillating toroidal dipoles. 
The linewidth of the collective mode [Fig.~\ref{fig:toroidal_lin}(c)] narrows strongly as the unit cell spacing decreases.

The transmitted light through the array can be calculated by adding the scattered light from each individual atom to the incoming light. At a positon $\zeta\unitvec{x}$ from a uniform lattice of area ${\cal A}$, when $\lambda\lesssim\zeta\ll \sqrt{{\cal A}}$ the electric field of the light transmitted in the forward direction is given by \cite{Chomaz12,Javanainen17,Facchinetti18,Javanainen19} 
\begin{equation}
\epsilon_0 \vec{E} = \epsilon_0\e_0\hat{\vec{e}}_y e^{ik\zeta} + {ik\over 2{\cal A}}\sum_j \left[\vec{d}_j-\hat{\vec{e}}_x\cdot \vec{d}_j \hat{\vec{e}}_x\right]e^{ik(\xi-x_j)}.
\end{equation}
The transmission $T=|\vec{E}|^2/|\e_0|^2$ shown in Fig.~\ref{fig:toroidal_lin}(d) displays narrow Fano resonances at the frequencies of toroidal and electric dipole excitations. (We note that a second dip at $\Delta=-15\gamma$ is due to coupling to an unrelated electric quadrupole mode.)

\begin{figure}
\centering
		\includegraphics[width=1\columnwidth]{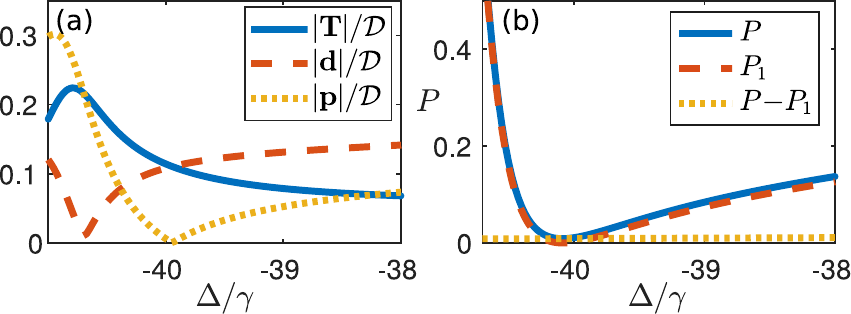}
    \vspace*{-12pt}
		\caption{ Excitation of a non-radiating dynamic anapole, with a multipole decomposition of (a) atomic dipoles [in units of $\mathcal{D}\mathcal{E}_0/(\hbar\gamma)$], with $\vec{p}=\vec{d}+ik\vec{T}$, and (b) the far-field scattered light from all contributions, the total dipole component with intensity proportional to $|\vec{p}|^2$, and the sum of all other contributions (in units of $I_\mathrm{in}/k^2$) as a function of the overall laser frequency detuning for $a=0.08\lambda$, $r=0.2\lambda$.  
		}
		\label{fig:anapole}
\end{figure}

An especially fascinating configuration can be obtained by a combination of a toroidal and electric dipole forming a dynamic anapole.
Because the far-field radiation of these dipoles is identical, they can destructively interfere such that the net radiation vanishes when $\vec{d}=-ik\vec{T}$.  Despite having no emission, the anapole state has a non-zero energy, and a vector potential which cannot be fully eliminated by gauge transformation~\cite{Afanasiev95,Nemkov17}.
We show that the collective radiative excitations of strongly coupled atoms can form a dynamic anapole by adding a pair of atoms to the toroidal dipole configuration of Fig.~\ref{fig:geometry}(a), in the same bilayer planes of the existing atoms, that then synthesizes a coherent superposition
of electric and toroidal dipoles [Fig.~\ref{fig:geometry}(b)]. The inset illustrates how the contribution of the electric dipole moment at the origin to the total dipole moment $\vec{p}$ points in the opposite direction to that of the toroidal dipole. 

Again, we illustrate the case of four squares, distributed evenly on a ring around the center but similar results can be achieved with a minimum of only two. As illustrated in Fig.~\ref{fig:geo_supp}(c), adding two central atoms at $\pm(a/2)\unitvec{x}$, results in an anapole excitation while all atoms lie in the $xy$ plane~\cite{SOM}.  

The anapole can be excited by a radially polarized plane-wave focused through a lens with high numerical aperture, leading to a longitudinal field in the $x$ direction along the beam axis which excites the central two atoms~\cite{SOM}. This field is calculated via the standard Richard-Wold diffraction integral~\cite{Richards59} with a numerical aperture of $0.7$. The resulting multipole decomposition of atomic dipoles [Fig.~\ref{fig:anapole}(a)] displays a strong excitement of both the electric and toroidal dipole. However, the combination of these dipoles, $\vec{p}=\vec{d}+ik\vec{T}$, is much weaker. The total scattered intensity, along with the decomposition into the dipole contribution and that of all other multipoles, is shown in Fig.~\ref{fig:anapole}(b), indicating a near total cancellation of the scattered light.

In conclusion, we have shown how strong light-mediated dipolar interactions between atoms can be harnessed to engineer collective radiative excitations that synthesize an effective dynamic toroidal dipole or anapole. 
In both cases the toroidal topology is generated by radiative transitions forming an effective poloidal electric current wound around a torus. In a large lattice we show how to engineer a collective strongly
subradiant eigenmode consisting of an effective periodic lattice of toroidal dipoles that exhibits a narrow
Fano transmission resonance.

 \begin{acknowledgments}
We acknowledge financial support from  EPSRC.
\end{acknowledgments}


\begin{thebibliography}{82}
\bibitem [{foo()}]{foot}%
  \BibitemOpen
  \href@noop {} {}\bibinfo {note} {The light and atomic polarization amplitudes
  here refer to the slowly varying positive frequency components, where the
  rapid variations $\exp(-i\omega t)$ at the laser frequency have been factored
  out.}\BibitemShut {Stop}%
\end{thebibliography}
\end{document}


\title{Supplemental Material to \\``Radiative toroidal dipole and anapole excitations in collectively responding arrays of atoms''}

\author{K.~E.~Ballantine}
\author{J.~Ruostekoski}

\affiliation{Department of Physics, Lancaster University, Lancaster, LA1 4YB, United Kingdom}

\date{\today}

\maketitle


\renewcommand{\thesection}{S\Roman{section}}
\renewcommand{\thesubsection}{S\Roman{section}.\Alph{subsection}}
\renewcommand{\theequation}{S\arabic{equation}}
\renewcommand{\thefigure}{S\arabic{figure}}
\renewcommand{\bibnumfmt}[1]{[S#1]}
\renewcommand{\citenumfont}[1]{S#1}

We include additional details of how the formalism employed in the main text for atomic transitions relates to effective oscillating currents and nanoresonator systems. We present a brief description of the mathematical model which applies both in the limit of low light intensity with coherent drive and to a single photon excitation in the absence of drive. We extend the discussion of the toroidal dipole eigenmode to the occupation of the mode. Additional details of how the proposal could be realized experimentally are considered. Finally we give results for simplified toroidal dipole and anapole unit cells with atoms only in a single 2D plane.

\section{Light-atom interactions}

\subsection{Quantum system of atoms and light}

In the main text we describe the excitation of an effective toroidal dipole and anapole in a coherently driven atomic ensemble by the equations $\dot{\vec{b}} =i\mathcal{H}\vec{b}+\vec{f}$ where $\vec{b}$ is a vector of atomic dipole amplitudes and $\vec{f}$ is proportional to the amplitude of the driving field. Here we describe the quantum model of atoms interacting with light in the \emph{length gauge}, obtained by the Power-Zienau-Woolley transformation~\cite{PowerZienauPTRS1959,Woolley1971a,CohenT,Ruostekoski1997a}. 
In the limit of low light intensity, the system with a single electronic ground state can be exactly described by a classical model of coupled dipoles driven by coherent light with field $\cbE(\vec{r})$~\cite{Javanainen1999a,Lee16}. 
The full many-excitation dynamics is described by the many-body quantum master equation for the density matrix~\cite{Lehmberg1970};
 \begin{equation}
\begin{multlined}
\label{eq:rhoeom}
\dot{\rho} = i\sum_{j,\nu}\left[H_{j\nu},\rho\right] +i\sum_{jl\nu\mu (l\neq j)}\Omega^{(jl)}_{\nu\mu}\left[\hat{\sigma}_{j\nu}^{+}\hat{\sigma}_{l\mu}^{-},\rho\right] \\
+\sum_{jl\nu\mu}\gamma^{(jl)}_{\nu\mu}\left(
2\hat{\sigma}^{-}_{l\mu}\rho\hat{\sigma}^{+}_{j\nu}-\hat{\sigma}_{j\nu}^{+}\hat{\sigma}_{l\mu}^{-}\rho -\rho\hat{\sigma}_{j\nu}^{+}\hat{\sigma}_{l\mu}^{-}\right) \,,
\end{multlined}
\end{equation}
with the square brackets representing commutators and
\begin{equation}
\begin{multlined}
H_{j\nu}=\Delta_{\nu}^{(j)}\hat{\sigma}_{j\nu}^{+}\hat{\sigma}_{j\nu}^{-} \\ + \frac{\xi}{D}\unitvec{e}_\nu\cdot \epsilon_0\mathcal{E}_0(\vec{r}_j) \hat{\sigma}_{j\nu}^{+} + \frac{\xi}{D}\unitvec{e}_{\nu}^{\ast}\cdot\epsilon_0\mathcal{E}_{0}^{\ast}(\vec{r}_j)\hat{\sigma}_{j\nu}^{-},
\end{multlined}
\end{equation}
where $\sigma_{j\nu}^{+}=|e\rangle_{j\nu}\mbox{}_{j}\langle g|$ is the raising operator for atom $j$ from its ground state to its excited state with $m=\nu$. 
Here, $\Delta_{\nu}^{(j)}$ is the detuning of each level from resonance. The Hermitian interaction terms $\Omega_{\nu\mu}^{(jl)}$ and collective dissipation terms $\gamma_{\nu\mu}^{(jl)}$ due to the dipole-dipole coupling between the atoms are given by the real and imaginary part of
\begin{equation}
\Omega^{(jl)}_{\nu\mu}+i\gamma^{(jl)}_{\nu\mu}=
\xi \mathcal{G}^{(jl)}_{\nu\mu} \, ,
\end{equation}
where $\mathcal{G}^{(jk)}_{\nu\mu}=\unitvec{e}_{\nu}\cdot\mathsf{G}(\vec{r}_i-\vec{r}_j)\unitvec{e}_{\mu}$ is the dipole radiation kernel that, acting on a dipole $\mathbf{d}$ at the origin, produces the familiar expression~\cite{Jackson} (with $\hat{\mathbf{r}}=\mathbf{r}/|\mathbf{r}|$)
\begin{align}\label{Gdef}
\mathsf{G}(\mathbf{r})\mathbf{d}&=-\frac{\mathbf{d}\delta(\mathbf{r})}{3}+\frac{k^3}{4\pi}\Bigg\{\left(\hat{\mathbf{r}}\times\mathbf{d}\right)\times\hat{\mathbf{r}}\frac{e^{ikr}}{kr}\nonumber\\
&\phantom{==}-\left[3\hat{\mathbf{r}}\left(\hat{\mathbf{r}}\cdot\mathbf{d}\right)-\mathbf{d}\right]\left[\frac{i}{(kr)^2}-\frac{1}{(kr)^3}\right]e^{ikr}\Bigg\},
\end{align}
and $\xi=6\pi\gamma/k^3$.

\subsection{Low light intensity limit}
\label{sec:LLI}

In the limit of low drive intensity we neglect terms containing products of two or more excited state amplitudes, or one or more excited state amplitudes  multiplied by the drive field~\cite{Ruostekoski1997a}. 
In our system, this amounts to neglecting terms $\langle\sigma_{j\nu}^{+}\sigma_{\ell\mu}^{-}\rangle$ (along with higher order correlators of $\sigma_{j\nu}^{\pm}$) and $\mathcal{E}_0\left<\sigma_m^+\right>$. 
The excited state population is negligible while the ground state population is invariant. The only non-trivial equation is for $\langle\sigma_{j\nu}^{-}\rangle$ and Eq.~\ref{eq:rhoeom} reduces to
\begin{equation}
\dot{\vec{b}}=i\mathcal{H}\vec{b}+\vec{f},
\end{equation}
in terms of the vector $b_{3j+\nu-1}=\langle\sigma_{j\nu}^{-}\rangle$, the matrix 
\begin{multline}
\mathcal{H}_{3j+\nu-1,3k+\mu-1} = \Delta_{\nu}^{(j)}\delta_{\nu\mu}\delta_{jk} + \Omega_{\nu\mu}^{(jk)}(1-\delta_{jk})+i\gamma_{\nu\mu}^{(jk)},
\end{multline}
and the drive $f_{3j+\nu-1} = i(\xi/\mathcal{D})\unitvec{e}_\nu\cdot\epsilon\cbE(\vec{r}_j)$.

\subsection{Single photon excitation}

The dynamics of a single photon excitation, decaying in the absence of drive, can also be described by the same formalism. Without drive, there is decay from the single-excitation state to the ground state, but the remaining single-excitation state remains pure, and the dynamics is linear in the amplitude of this state. It is this linear response without saturation which provides identical single-particle amplitudes to those of the low light intensity limit of coherently driven atoms~\cite{SVI10}.

We again start from the full atomic equations of motion Eq.~\ref{eq:rhoeom}, but now assume that the initial state consists of a pure, single-photon excitation. Then the density matrix at later times can be written as
\begin{equation}
\rho = \ket{\Psi}\bra{\Psi} + p_g\ket{G}\bra{G},
\end{equation}
where $\ket{\Psi}$ is a state consisting of exactly one excitation, $\ket{G}$ is the state with all atoms in their ground state, and $p_g$ is the probability that the excitation has decayed. In this case there is no incoherent mixing between the excited and ground state.
While dissipation means that the norm of $\ket{\Psi}$ is not conserved, the dynamics within the single-excitation subspace is coherent, and it can always be expanded in terms of the individual atomic excitations
\begin{equation}
\ket{\Psi}= \sum_{j,\nu} \Pc^{(j)}_{\nu}(t)\,\hat{\sigma}^{+}_{j\nu}\ket{G},
\end{equation}
with amplitudes $\mathcal{P}_\nu^{(j)}(t)$. For single-particle expectation values, the dynamics can equally be written in terms of these amplitudes. In terms of the vector $b_{3j+\nu-1}=\mathcal{P}_\nu^{(j)}$, we have again $\dot{\vec{b}}=i\mathcal{H}\vec{b}$~\cite{Ballantine20ant}, formally equivalent to the equations describing the low light intensity limit in the absence of drive. The identical description is due to the fact that in each case the relevant dynamics includes only a part of the density matrix which evolves linearly without saturation, with the collective response being determined by the dipole propagation kernel $\mathcal{G}^{(jl)}_{\nu\mu}$.

\section{Collective toroidal eigenmode}
\label{sec:eigenmode}

\begin{figure}
\centering
		\includegraphics[width=1\columnwidth]{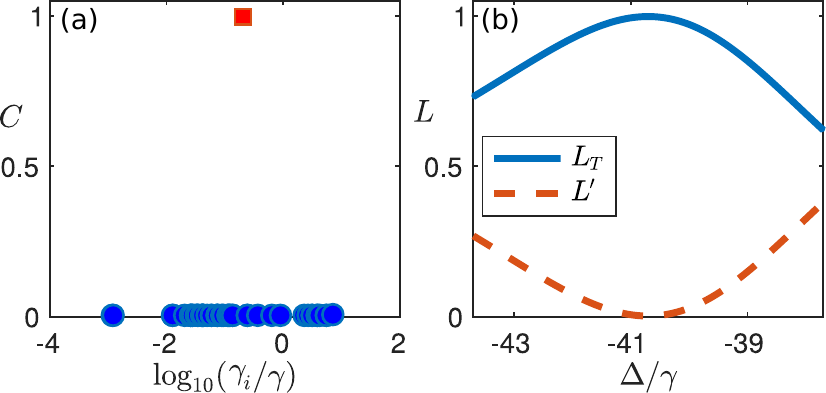}
    \vspace*{-12pt}
		\caption{
		(a) Overlap of each eigenmode of a toroidal dipole unit cell (ordered by the collective resonance linewidth) with an ideal toroidal dipole mode with poloidal polarization.
		(b) Occupation measure $L_T$ of the toroidal dipole collective eigenmode and sum $L^\prime$ of the occupation of all other modes in the steady state response, as laser detuning is varied, for $r=0.2\lambda$, $a=0.08\lambda$.  
		}
		\label{fig:L_supp}
		\end{figure}

The collective dynamics of the atomic ensemble is determined by the eigenvectors $\vec{v}_j$ of $\mathcal{H}$, representing the collective radiative excitation eigenmodes of the atoms, and the corresponding eigenvalues $\delta_j+i\gamma_j$~\cite{Jenkins_long16}, where $\gamma_j$ denotes the collective resonance linewidth and $\delta_j$ the collective resonance line shift from the single-atom resonance. While these eigenvectors are not orthogonal in general, they do form a basis, and the state can be expanded at all times as $\vec{b}(t)=\sum_i c_i(t) \vec{v}_i$. The occupation of each eigenmode can then be defined as~\cite{Facchinetti16}
\begin{equation}
L_i = \frac{|\vec{v}_i^T\vec{b}|^2}{\sum_j|\vec{v}_j^T\vec{b}|^2}.
\end{equation}
For a single-photon excitation, each eigenmode will decay individually with
\begin{equation}
c_j(t) = \exp{[(i\delta_j-\gamma_j)t]},
\end{equation}
with $c_j(0)$ determined by the initial state.

The toroidal dipole unit cell has a collective eigenmode exhibiting a strong toroidal dipole as discussed in the main text. We can similarly measure the overlap of this mode with an ideal toroidal dipole eigenmode $\vec{v}_T$, depicted in Fig.~1 of the main text, with poloidal polarization. This is given by $C_i=|\vec{v}_i^T\vec{v}_T|^2/\sum_j|\vec{v}_j^T\vec{v}_T|^2$. This overlap is illustrated for each eigenmode in Fig.~\ref{fig:L_supp}(a) showing that there is a unique collective mode which approximates the ideal poloidal polarization mode. The occupation $L_T$ of this eigenmode, as well as the sum of occupations of all other modes, is plotted in Fig.~\ref{fig:L_supp}(b) as a function of laser detuning, with $L_T>0.99$ at resonance. The maximum occupation in Fig.~\ref{fig:L_supp} corresponds to the maximum toroidal dipole excitation of Fig.~3 in the main text.

\begin{figure}
\centering
		\includegraphics[width=1\columnwidth]{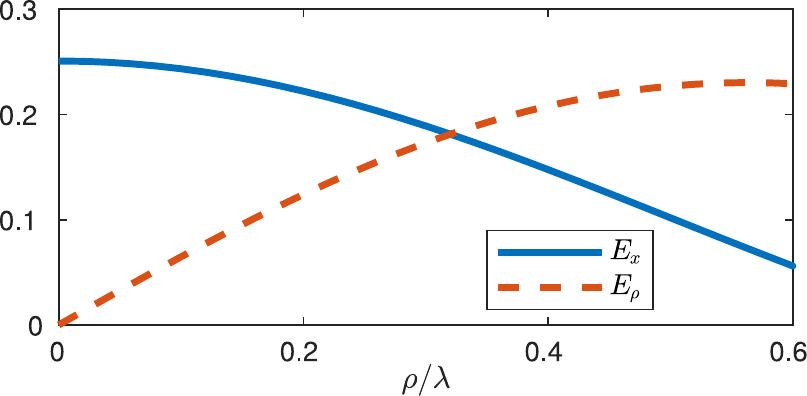}
    \vspace*{-12pt}
		\caption{
		Longitudinal component $|E_x|$ and radial component $|E_\rho|$ of the focused radially polarized beam used to excite the anapole, as a function of the distance $\rho=\sqrt{y^2+z^2}$ from the beam axis in the focal plane $x=0$. Here the incoming field before focusing is $\cbE=\unitvec{e}_\rho$ and the NA is $0.7$. 
		}
		\label{fig:components}
		\end{figure}

\section{Classical analogue of polarization distribution}

In this work, we have synthesized collective radiative excitations of atoms that produce a toroidal dipole and an anapole. This is achieved by a simple arrangement of atoms that experience strong light-mediated interactions.
In particular, the toroidal dipole is generated by the orientation of the atomic transitions due to the dipole-dipole interactions forming an effective poloidal electric current wound around a torus. Each atom produces an effective electric dipole generated by quantum-mechanical
electric-dipole-allowed transitions in electronic orbitals. 
Here we provide a classical analogue description of this quantum-mechanical process, illustrating how effective polarization and current densities originate from such atomic dipole transitions. This allows us to make comparisons with
systems of nanoparticles and solid-state artificial metamaterial resonators that have been used in studies of electromagnetic multipole radiation. 

\subsection{Classical effective currents from trapped atoms}

Classically, we describe the interactions of light with atoms trapped in a ring-shaped pattern, represented by a sequence of oscillating charges.
Consider a ring in the $xy$ plane with radius $\rho$, and let $\phi$ be the azimuthal angle in the plane. Suppose there is a flow of charge on the ring and consider an infinitesimal segment where a negative charge $-q$ moves from a point $\vec{r}_\phi$ at angle $\phi$  a distance $\Delta\vec{r}=-\rho\,\mathrm{d}\phi\unitvec{\phi}$ along the ring in a time $\mathrm{d}t$, leaving a positive charge $q$ stationary at $\vec{r}_\phi$. The charge distribution can be replaced by $n$ dipoles with spacing $\Delta r/n$ such that the positive charge of each dipole overlaps with the negative charge of the next, leaving only the original charges at each end. Then the polarization density in the limit $n\rightarrow \infty$ is~\cite{CohenT} 
\begin{equation}
\label{eq:pol}
\begin{multlined}
\vec{P}=\lim_{n\to \infty}\sum_{p=0}^{n-1} (-q)\,\frac{\Delta\vec{r}}{n}\delta\left(\vec{r}-\left(\vec{r}_\phi+\frac{p+1/2}{n}\right)\Delta\vec{r}\right) \\
= -\int_{0}^{1}\mathrm{d}u\, q\Delta\vec{r} \delta(\vec{r}-\vec{r}_\phi-u\Delta\vec{r}),
\end{multlined}
\end{equation}
giving a continuous polarization density consisting of point dipoles $\mathrm{d}\vec{P}=-q\Delta\vec{r}\delta(\vec{r}-\vec{r}_\phi-u\Delta\vec{r})\mathrm{d}u=q\rho\,\mathrm{d}\phi\,\delta(\vec{r}-\vec{r}_\phi-u\Delta\vec{r})\mathrm{d}u\unitvec{\phi}$ located at $\vec{r}_\phi+u\Delta\vec{r}$ and pointing in the direction $\unitvec{\phi}$. These point dipoles describe the response of single atoms, whose size is negligible compared to the optical wavelength.

The current density associated with the change in the electric charge distribution from each point dipole is given by
\begin{align}
\mathrm{d}\vec{J} &= \frac{1}{\mathrm{d}t}[-q(\vec{r}_\phi+\Delta\vec{r}_\phi)-(-q)\vec{r}_\phi]\delta(\vec{r}-\vec{r}_\phi-u\Delta\vec{r})\mathrm{d}u,\\
&=q\rho\frac{\mathrm{d}\phi}{\mathrm{d}t}\delta(\vec{r}-\vec{r}_\phi-u\Delta\vec{r})\mathrm{d}u\unitvec{\phi},\nonumber
\end{align}
which is simply equal to $\mathrm{d}\vec{P}/\mathrm{d}t$. This circulating current, for a sufficiently small circle, creates an effective magnetic dipole with a magnetic moment $d\mu =\mathrm{d}u (q/(2\mathrm{d}t))\vec{r}_\phi\times\Delta\vec{r}=\mathrm{d}u(q \rho^2\mathrm{d}\phi/(2\mathrm{d}t))\unitvec{z}$ such that the total magnetic moment $\mu=\int \mathrm{d}\mu = (\mathrm{d}\phi \rho^2/2) (q/\mathrm{d}t) \unitvec{z}$ is simply the current times the area of the circle within the angle $\mathrm{d}\phi$. 


While this description applies to a continuous distribution of dipoles located at $\vec{r}_\phi+u\Delta\vec{r}$ (or $n$ discreet dipoles without taking the limit $n\rightarrow \infty$ in Eq.~\ref{eq:pol}), it can be approximated with a discrete series of a smaller number of dipoles located at fixed positions $\vec{r}_j$. The corresponding polarization density $P_\sigma(\vec{r},t)=\exp{(-i\omega t)}\mathcal{D}\sum_j \mathcal{P}_\sigma^{(j)}\delta(\vec{r}-\vec{r}_j)$ induced by a driving field with frequency $\omega$ leads to a current density $\vec{J}=-i\omega \vec{P}$. When these discrete currents are arranged to point tangentially to a circle, they approximate a closed loop of current. The magnetic dipole moment appearing in the scattering cross section is then~\cite{Alaee18} $\vec{m}=\mu/c=1/(2c)\int \mathrm{d}^3\vec{r}\, \vec{r}\times\vec{J}(\vec{r})$. 
Similarly, poloidal current distributions are approximated by currents from time dependent point dipoles giving a toroidal dipole moment~\cite{Papasimakis16}
\begin{equation}
\vec{T}=\frac{1}{10 c}\int \mathrm{d}^3\vec{r}\left[\vec{r}(\vec{r}\cdot\vec{J})-2\vec{r}^2 \vec{J}\right].
\end{equation}

Numerically, we include all the light-mediated dipole-dipole interactions between the atoms and calculate the collective excitation eigenmodes for the coupled dipoles (see Sec.~\ref{sec:eigenmode}). One of the eigenmodes 
then corresponds to the poloidal current configuration of the toroidal dipole [Fig.~\ref{fig:L_supp}(a)]. This also applies to the case of an array of several toroidal dipoles, as discussed in the main text.

\subsection{Nanoparticles and solid-state resonators}

The classical description of the effective current due to atomic dipole transitions, given in the previous section, allows us to make comparisons with systems of nanoparticles and resonators in artificial metamaterials that
form multipole radiation sources.
The most dramatic difference is the quantum-mechanical nature of the optical interaction for the case of atoms, where the atomic transitions are determined by the precise resonance frequency and the quantum-mechanical Wigner-Weisskopf resonance linewidth~\cite{CohenT}. Atoms also form truly pointlike electric dipoles.

In the limit of low light intensity, the atoms interacting with incident light can be considered as a linear classical coupled-dipole system (see Sec.~\ref{sec:LLI}).
Small nanoparticles~\cite{Alu08,Evlyukhin10} or circuit resonators~\cite{JenkinsLongPRB,CAIT} are frequently approximated as effective coupled point dipoles in electromagnetic fields. Nanoparticles with their electric dipoles (provided that the higher-order multipole contributions are negligible) forming a closed ring can then exhibit a mode that behaves as
an effective magnetic dipole~\cite{Alu08}, analogously to the earlier classical charge distribution description of the atoms. In resonant $LCR$ circuits in metamaterials, the current oscillations form effective electric and magnetic dipoles that radiatively couple with the current oscillations of the other circuits. If each circuit is modeled as a pointlike particle, and higher-order multipole contributions of each circuit are negligible, the dynamics corresponds to a coupled-dipole system~\cite{JenkinsLongPRB,CAIT}.

As a comparative example, a nanorod can be approximated as an effective point dipole~\cite{Watson2019}, where the polarization density is to first order assumed to be a uniform distribution of point dipoles throughout the volume of the rod. Depending on the 
geometry of the problem and the thickness of the nanorods, all other excitation modes, such as radial or azimuthal currents are then in this approximation ignored. 
For a single nanorod of radius $a$ and length $H$ at the origin, and aligned along the $z$ axis, the resulting polarization distribution is 
\begin{equation}
\vec{P} = \frac{\mathcal{Q}(t)}{\pi a^2}\unitvec{z}\Theta(a-\rho)\Theta(H/2-z)\Theta(H/2+z),
\end{equation} 
where $\Theta$ is the Heaviside function, $\rho=\sqrt{x^2+y^2}$, and $\mathcal{Q}(t)$ is a generalized coordinate whose derivative $\dot{\mathcal{Q}}(t)$ describes current oscillations in the nanorod. In contrast to atomic dipoles, although for $H\alt \lambda$ this polarization density can be approximated by a point dipole $\vec{P}=\mathcal{Q}(t) H \unitvec{z}\delta(\vec{r})$, this approximation easily breaks down for interacting nanorods that are too closely separated compared with the resonant wavelength~\cite{Watson2019}.

\section{Experimental considerations}

For most of the examples discussed in the main text we choose $r=(n+1/2)a$ for integer $n$, with the result that the atoms lie on selected sites of a square lattice with lattice constant $a$ (see main text Fig.~2). Four intersecting beams, two pairs of propagating beams in the $y$ and $z$ directions respectively, can be used to produce such a lattice with confining potential 
\begin{equation}
V(\vec{r}) = sE_R\left[\sin^2{\left(\pi\frac{y}{a}\right)}+\sin^2{\left(\pi\frac{z}{a}\right)}\right],
\end{equation}
where $E_R=\pi^2\hbar^2/(2ma^2)$ is the lattice recoil energy and $s$ is a dimensionless constant which determines the depth of the lattice~\cite{Morsch06}. An additional potential can confine the atoms in the $x=\pm a/2$ planes. This could be an identical pair of counter-propagating beams in the $x$ direction, or a simpler double-well potential. 
Locally, each atom experiences a harmonic trapping potential $V(\vec{r})=(m/2)\sum\omega_\mu^2(\Delta r_\mu)^2$ where $\omega_{y,z}=2\sqrt{s}E_R/\hbar$, $\omega_x$ is the harmonic trapping frequency in the $x$ direction, and $\Delta \vec{r}$ is the displacement from the center of the lattice site. Then the atom at site $\vec{r}_j$ has a Wannier function $\phi_j(\vec{r})=\phi(\vec{r}-\vec{r}_j)$ given by
\begin{equation}
\phi(\vec{r}) = \frac{1}{(\pi^3l^4l_x^2)^{1/4}}\exp{\left(-\frac{y^2+z^2}{2l^2}-\frac{x^2}{2l_x^2}\right)},
\end{equation}
with width $l=as^{-1/4}/\pi$ and thickness $l_x=\sqrt{\hbar/(m\omega_x)}$. The atoms can be increasingly strongly confined by increasing the trapping strength $s$. Experimentally, atoms are loaded deterministically into a Mott-insulator state with one atom per site~\cite{Schnorrberger09,ShersonEtAlNature2010,Rui2020}, and the desired geometry achieved by removing excess atoms on a site-by-site basis~\cite{Weitenberg11}.

Further techniques have also been developed to provide deeply subwavelength features in optical trapping potentials. For example, coupling three atoms in a $\Lambda$ configuration via a strong control field with Rabi frequency $\Omega_c(y,z)=\Omega_c\sin{(ky)}\sin{(kz)}$ and a weak probe field with Rabi frequency $\Omega_p$ leads to a trapping potential which depends on the ratio $\Omega_c(y,z)/\Omega_p$, which varies rapidly in a small region close to the nodes of $\Omega_c(y,z)$, for $\Omega_p/\Omega_c\ll 1$~\cite{Wang18}. Alternatively, internal degrees of freedom can be exploited~\cite{Anderson20}.  Lattices with periodicity less than half the wavelength of the control field can also be engineered by stroboscopicaly shifting the lattice at high frequency such that the atoms experience a time-averaged potential with higher periodicity than the instantaneous potential at any one time~\cite{Nascimbene15,Lacki19}.  Finally, optical tweezers provide an alternative means to design arbitrary potentials~\cite{Xia15,Lester15,Cooper18} with single-site control~\cite{Endres16,Barredo1021}.

Atoms such as Sr and Yb are particularly suitable for subwavelength trapping. $^{33}$Sr has a transition between the $^3P_0$ state and the triply degenerate $^3D_1$ state~\cite{Olmos13} with wavelength $\lambda=2.6\mu$m and linewidth $\Gamma=2.9\times10^{5}$/s. The magic wavelength for these states gives an optical lattice with spacing $d=206.4$nm, less than $\lambda/10$. While most of the examples we consider require varying only the overall laser detuning, the data shown in Fig.~4 of the main text requires the individual atomic level shifts of different atoms to be controlled. This could be achieved by ac Stark shifts~\cite{gerbier_pra_2006} from standing wave lasers offset from the trapping lattice such that different sites experience different intensities.

\section{Anapole excitation}

To excite the anapole we use a tightly focused radially polarized beam. The focusing leads to a longitudinal component in the $x$ direction on the beam axis which directly drives the $x$ component of polarization on the two atoms at the center of the anapole. Off axis, a combination of radial and longitudinal polarization couple to the toroidal dipole mode of the remaining atoms. The amplitude of the longitudinal and radial components of the field in the $x=0$ focal plane are shown in Fig.~\ref{fig:components} as a function of the radial distance $\rho$ from the beam axis.

\section{In-plane toroidal dipole and anapole}

\begin{figure}
\centering
		\includegraphics[width=1\columnwidth]{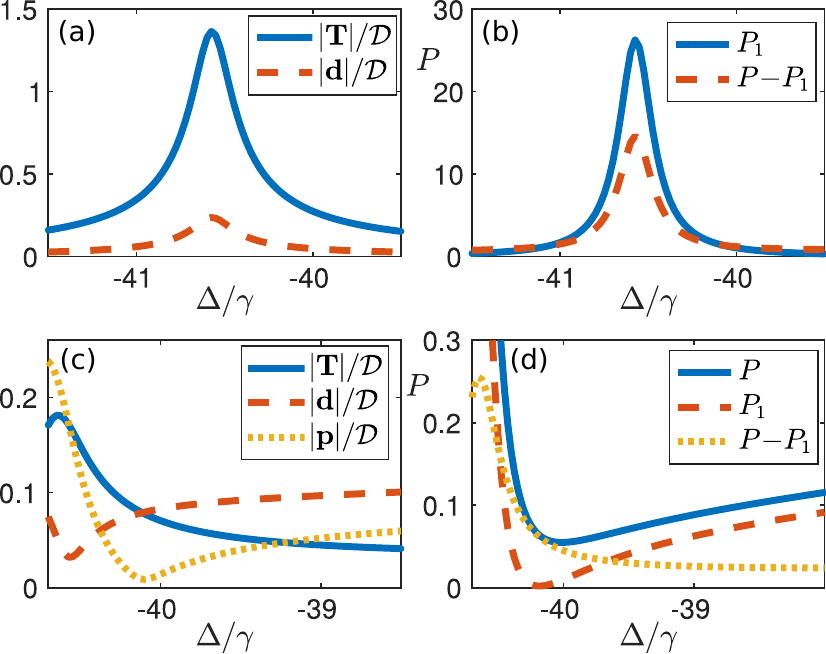}
    \vspace*{-12pt}
		\caption{
		Excitation of toroidal dipole and dynamic anapole with two magnetic dipole moments such that all atoms are in the single $xy$ plane. Multipole decomposition for toroidal dipole unit cell of (a) atomic dipoles [in units of $\mathcal{D}\mathcal{E}_0/(\hbar\gamma)$], and (b) the far-field scattered light, separated into the total dipole component, and the remaining sum of all other contributions (in units of incident light intensity $I_{\mathrm{in}}/k^2$). Multipole decomposition for the single-plane dynamic anapole of (c) atomic dipoles [in units of $\mathcal{D}\mathcal{E}_0/(\hbar\gamma)$], with $\vec{p}=\vec{d}+ik\vec{T}$, and (d) the far-field scattered light, separated into the total dipole component, and the remaining sum all other contributions (in units of $I_{\mathrm{in}}/k^2$). In all cases  $r=0.2\lambda$ and $a=0.08\lambda$.
		}
		\label{fig:anapole_supp}
		\end{figure}

An experimentally even simpler realization of both the toroidal dipole and the anapole with atoms confined only to a single plane can be achieved by removing the atoms for which $z\neq 0$, as shown in Fig.~2(c) of the main text, leaving two squares with opposite chirality polarization. These squares will have magnetic dipole moments pointing in the $\unitvec{z}$ and $-\unitvec{z}$ directions, respectively, leading to zero net magnetic moment but contributing to a toroidal dipole. The resulting multipole decomposition of the atomic dipoles and the far-field scattered light is shown in Fig.~\ref{fig:anapole_supp}(a,b). A strong toroidal dipole is present, although the far-field scattered light in this case also shows a comparable contribution from higher order modes. Again this toroidal dipole can interfere with a net electric dipole moment on two atoms at $\pm(a/2)\unitvec{x}$ to form an anapole. The resulting multipole decomposition is shown in Fig.~\ref{fig:anapole_supp}(c,d). There is strong suppression of the total dipole moment and a significant dip in the scattered light, although again the contributions of other multipole moments, especially quadrupoles, is stronger than the case with four squares presented in the main text. 
These contributions of higher-order multipole moments could be further suppressed by adding more atomic squares, 8, 16, etc.

%